# A spiral sampling method for calculating the complex orbital angular momentum spectrum

Zheng Han,[†] Bowen Yang,[†] Xiao Chen,[*] Yiquan Wang

*College of Science, MINZU University of China, Beijing 100081, China*
*†The authors contributed equally to this work.*

**OAM spectrum reflects the OAM component included in measured light field which is crucial in OAM -based application. However, traditional definition-based OAM spectrum algorithm is extraordinary time-consuming and limited to prior knowledge severely. To overcome it, in this paper, we propose a novel method to calculate the complex spectrum. After spiral sampling and Fourier transform, one can retrieve the radial coefficient of arbitrary OAM component by proper filtering and inverse Fourier transform. The simulation results reveal that the Mean Absolute Error (MAE) between retrieved one and target could reach at 0.0221 and 0.0199 on average for amplitude and phase respectively after normalized. This method could provide a powerful tool for future OAM-based and application.**

Since the discovery of orbital angular momentum (OAM) carried by photons in 1992, light beams carrying OAMs have been utilized widely in numerous research fields including optical communication, holography, optical tweezers and quantum information[1]. Among those applications, a knowledge about OAM component contained in the light field is crucial, which could help us understand the nature of field better. Based on that, the concept of OAM spectrum was proposed. Due to the orthogonality of the helical phase factor $e^{il\theta}$ with different topological charge *l*, a superimposed vortex light field could be decomposed into a series of simple field containing only a kind of OAM, hence we can view the complex field more concisely. In analogy with optical spectrum, a OAM spectrum could provide both amplitude and phase information with respect to different topological charges in transverse light field. As a result, a comprehensive and efficient analysis of OAM beams is crucial, which could be seen as a powerful tool in various scenarios.

So far, a lot of methods have been proposed on this task. For example, rotational Doppler frequency shift[2], spatial mode decomposition[3], Kramers–Kronig retrieval method[4], single-pixel detector method[5, 6] could be utilized to achieve both the amplitude and phase retrieval. What's more, as the development of deep learning, training an hybrid opto-electronic neural network[7] to fit the relationship between the input OAM light field (or intensity only) and output spectrum index have been verified feasible.

Although amazing progress has been obtained, most of present researches focus on the innovation of collecting light field or direct measurement through physical devices, but ignoring the improvement of algorithm. Actually, traditional definition-based OAM spectrum algorithm is not only extraordinary time-consuming, but also forcing researchers must select the measured topological charge range manually before starting calculation, resulting in the loss of information included in data more or less due to an empirical operation.

In response to the issue, we propose a novel OAM spectrum algorithm based on spiral sampling and angular Fourier transform in this article, which could accelerate the computational procedure and extend the maximal measurable topological charge to half of the angular sampling frequency simultaneously. Inspired by[8], by exploiting similarity between the helical phase factor $e^{il\theta}$ and exponential frequency factor $e^{i\omega t}$, we can retrieve OAM spectrum through the angular Fourier transform. More than that, in order to obtain OAM spectrum of the whole plane at once, we adopt an Archimedes' spiral to sample the light field, which could break the restriction of one circle sampling. After Fourier transforming to the sampling data, we could extract desired complex radial coefficient by applying a corresponding filter and implementing inverse Fourier transform. The simulation results validate the effectiveness of our method, which will provide a useful tool in analyzing the component of OAM fields.

Generally, given a particular frequency and propagating distance, the transverse field of superimposed OAMs could expressed simply as:

$$E(r,\theta) = \sum_{l=N_1}^{N_2} a_l(r)e^{il\theta} \qquad (1)$$

As shown in eq. (1), unknown field could be extended into a series form of helical modes $e^{il\theta}$, *l* indicates the topological charge. $a_l(r)$ is the complex radial coefficient of a special helical modes, which is defined similar to Fourier series:

$$a_l(r) = \frac{1}{2\pi}\int_0^{2\pi} E(r,\theta)\, e^{-il\theta}d\theta \qquad (2)$$

The value of $a_l(r)$ characterizes the amplitude and phase distribution of corresponding single OAM field, therefore we will focus on the retrieval of it.

In order to obtain the information of every $a_l(r)$ with different *l* and *r*, instead of depending on definition based method to calculate all coefficient one by one according to eq. (2), we adopt a spiral sampling method to support faster and more general OAM spectrum retrieval, where the maximal OAM component that can be measured reaches to half of the angular sampling frequency. An Archimedes' spiral could be written in polar coordinate:

$$r = r_0 + \alpha\theta \qquad (3)$$

$(r, \theta)$ represents the polar coordinate, $r_0$ indicates the initial spiral length and $\alpha$ is a scaling factor. From eq. (3), we establish a correlation between mutually independent $r$ and $\theta$. For simplicity, we set $r_0 = 0$. If we use this spiral to sample the input superimposed OAM field, we could derive that:

$$E'(\theta) = \sum_{l=N_1}^{N_2} a_l(\alpha\theta) e^{il\theta} \quad (4)$$

Through this procedure, the 2 dimensional field data is converted to 1 dimension without losing any information about the $a_l(r)$. If we observe the form of eq. (4), it's very similar to traditional Frequency Division Multiplexing (FDB) signals with Double-Sideband Suppressed-Carrier (DSB-SC) modulation. The $a_l(\alpha\theta)$ in eq. (4) could be treated as the modulating signal and helical phase factor $e^{il\theta}$ could be seem as carrier if we replace the angular variable $\theta$ to time variable $t$ slightly. For our method to be consistent with Fourier transform theory, we call $e^{il\theta}$ as angular frequency factor in the following part of this article.

Next step, we take the Fourier transform of $E'(\theta)$. The angular frequency factor $e^{il\theta}$ will bring a frequency shift to the spectrum of corresponding $a_l(\alpha\theta)$. It could be written as:

$$F(j\iota) = \sum_{l=N_1}^{N_2} \frac{1}{|\alpha|} A\left[j\left(\frac{\iota}{\alpha} - l\right)\right] \quad (5)$$

Here, we use $\iota$ to represent the angular frequency variable and $A(j\iota)$ indicates the spectrum of $a(\theta)$. Eq. (5) shows the frequency domain expression of $E'(\theta)$. Sampled complex radial coefficients $a_l(\alpha\theta)$ for different OAM component will be separated in the frequency domain with the help of angular frequency factor $e^{il\theta}$ and only be added a scaling factor $\alpha$ compared with original ones.

To acquire arbitrary $a_l(\alpha\theta)$, we could multiple a carrier $e^{-il\theta}$ to $E'(\theta)$ before taking Fourier transform firstly and then apply an appropriate low-pass filter, further take inverse Fourier transform of it. In the end, we rescale this coefficient by multiple $\alpha^{-1}$ to angular variable $\theta$ to recover $a_l(r)$. It's worth noting that $\alpha$ is a crucial factor in this method because it could provide scaling to $A\left[j\left(\frac{\iota}{\alpha}\right)\right]$. To avoid the spectrum overlapping, the intercarrier distance must be larger than the half-bandwidth sum of adjacent signals. All above is summarized in Fig. 1.

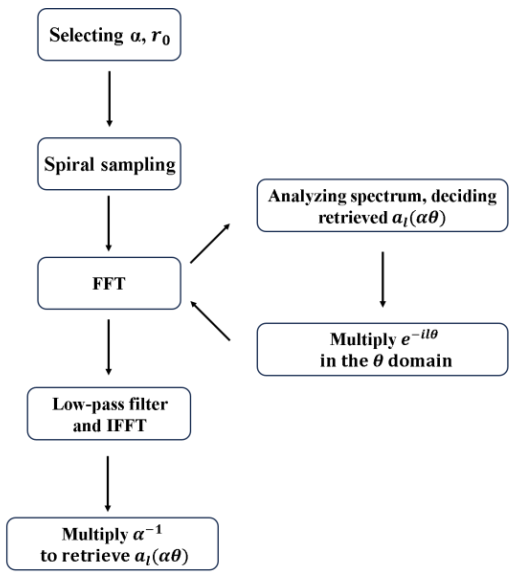

Fig. 1. Flow chart of spiral sampling method to calculate the OAM spectrum.

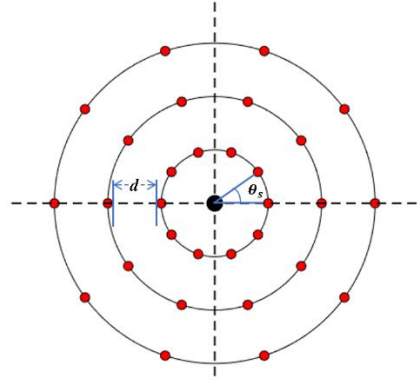

Fig. 2. Schematic diagram of spiral sampling in discrete domain.

In practice, because the computer could also operate discrete data, so we must consider discrete signal processing, so the spiral is transformed to a series of sampling point instead of a continuous curve. According to the Nyquist sampling theorem, the maximal signal frequency that can be reached is the half of sampling frequency, so the measurable maximal OAM component (topological charge) is extended to half of angular sampling frequency, which free from manual selection in traditional definition-based method. The complex radial coefficient $a_l(r)$ could be seen as slow-changing signals so it's very easy to fulfill the Nyquist sampling theorem in the process of radial sampling, we'll consider it to be sampled very adequately in the following analyzation. As shown in Fig. 2, now we define the angle between adjacent angular sampling points $\theta_s$ and radial sampling point $d$ because it's unchanged device parameter normally. It's economical to reconsider to set an appropriate bias $r_0$ because there's no intensity in the center of OAM beams. Then we define a parameter $\theta_n$ to characterize the angle that spiral go through when it extends a radial length $d$. The $\theta_n$ must be an integral multiple of $\theta_s$ like $\theta_n = n\theta_s$, and similarly $r_0 = nd, n = 1,2,3,....$

We could construct an equation of this spiral:

$$\begin{cases} r = r_0 + \alpha\theta & (6) \\ \theta = 0, r = r_0 & (7) \\ \theta = \theta_n, r = r_0 + d & (8) \end{cases}$$

Solve this equation, we could obtain:

$$\alpha = \frac{d}{\theta_n} = \frac{d}{n\theta_s} \quad (9)$$

It's shown that a larger $\theta_n$ will lead to smaller $\alpha$. choosing $\alpha$ less than one means to compress the spectrum range of original $a_l(r)$, which is beneficial to avoid spectrum overlapping. However, larger $\theta_n$ causes the reduction of angular sampling frequency, resulting in the information loss

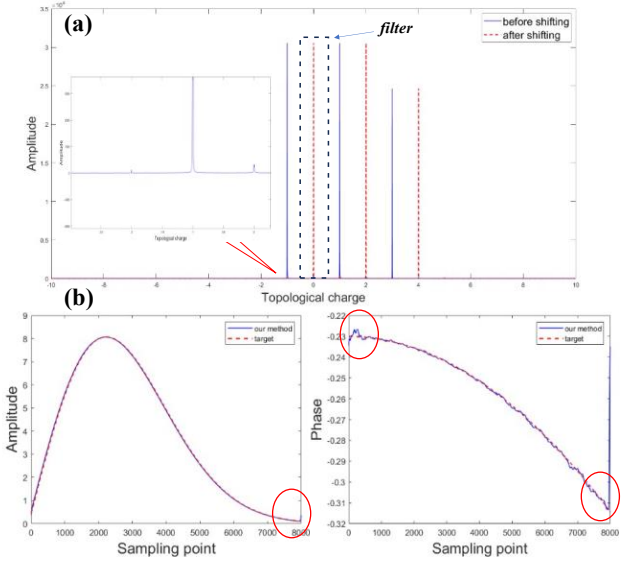

Fig. 3. (a) The spectrum of spiral sampling light field ($l$ = -1,1,3). The x-coordinate is topological charge and y-coordinate shows the amplitude of complex spectrum. The original one (solid blue) moves right (dash red) in order to filter the $l$ = -1 component. (b) The comparison of amplitude and phase distribution between filtered one and target. Red circles indicate the region that shows relative large deviations.

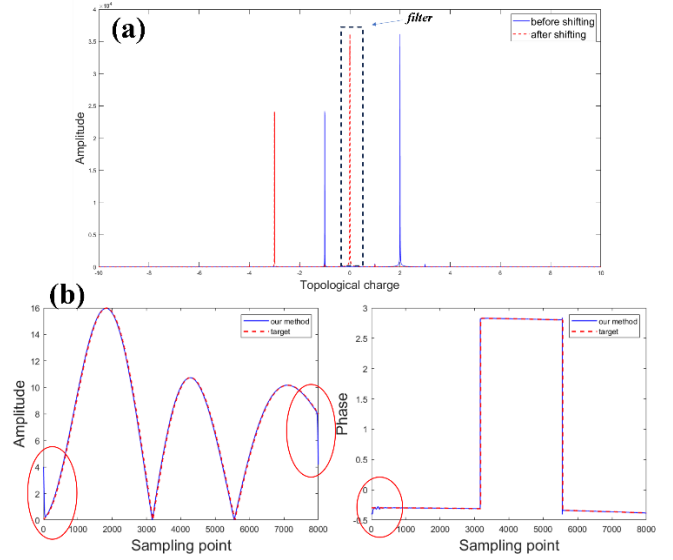

Fig. 4. (a)The spectrum of spiral sampled LG beams(p=1,l=-1),(p=2,l=2). The others correspond with Fig. 3.

of higher OAM components. It's a trade-off that we need to consider carefully in practice.

We validate the feasibility of our method in numerical simulation. Without loss of generality, there is a superimposed OAM field ($l$ = -1, 1, 3) shown in Fig. 2, here we show an example of retrieving $a_{l=-1}(r)$, we set $\alpha = 0.2$ empirically. In Fig. 2(b), the whole spectrum shifts from solid line part to dash line part in order to deploy the $l$ = -1 part into the baseband. After that we could retrieve $a_{l=-1}(r)$ by following the flow chart in Fig. 1. To evaluate the performance of our method, we normalize them and compute the Mean Absolute Error (MAE) to characterize the difference between calculated one and target. It's defined as:

$$MAE(y, \hat{y}) = \frac{1}{N} \sum_{i=1}^{N} |y_i - \hat{y}_i| \quad (10)$$

$y$ indicates the target and $\hat{y}$ is the result of our method. The filtered output accords with the target well where the MAPE of amplitude and phase are 0.011 and 0.0321 respectively. It's noted that the filter is needed to select desired spectrum component and introducing crosstalk as few as possible simultaneously. The bandwidth of $a_l(\alpha\theta)$ is up to the radial function and scaling factor $\alpha$, and the specific design of filter depends on the bandwidth of all $a_l(\alpha\theta)$ with different $l$. The relevant knowledge about digital filter and discrete signal processing could be found in Ref. [9].

Similarly, we also demonstrate the superposition of Laguerre-Gaussian (LG) beams with non-zero radial index $p$. The single LG beam with non-zero $p$ exhibits a multi-ring intensity structure, therefore the radial function will show different distribution compared with common vortex beams. In Fig. 3, we try to acquire the radial coefficient of $(p = 2, l = 2)$ among the superimposed beams including $\{(p = 1, l = -1), (p = 2, l = 2)\}$. The results in Fig. 4(b) reveals that we retrieve $a_{p=2,l=2}(r)$ successfully again with slight error, which the MAE are 0.0332 and 0.0077 respectively. It's noted that through the comparison between amplitude and phase distribution of recovered and target signals in both Fig. 3 and Fig. 4, we could observe the fluctuation in the boundary region of recovered signal marked in red circles, that's due to the boundary effect of filter, nevertheless it could be reduced further by reasonably choosing filters with higher peak sidelobe ratio (PSLR) such as Blackman Filter and adjusting parameters carefully. Or to say, if we only care about tendency but specific value, there will be no problem.

Finally, we'd like to talk about the influence of uneven sampling. As discussed above, both the OAM beams and spiral sampling method are based on polar coordinate, resulting in errors if we record light field data using commercial CCD device in Cartesian coordinate.

As illustrated in the insert of Fig. 3(a), there are undesired spectrum components distributed on both side of correct OAM component symmetrically, which could be seen as spectrum spreading, resulting in the interference to desired component. That's due to the uneven spiral sampling under Cartesian coordinate. As a result, it's better suitable to employ receiving device based on polar coordinate such as ring-shape photodetector array.

In this letter, we propose a novel spiral sampling method to calculate the complex OAM spectrum, comparing with existing scheme, ours could extend the maximal measurable topological charge to half of the angular sampling frequency, leading to the liberation from manual selection, moreover, the computing consumption decreased simultaneously thanks to the FFT algorithm. This method could provide a powerful tool for future OAM-based application such as optical communication and holography.

**Funding.**

**References**

1. Y. Shen, X. Wang, Z. Xie et al., Light: Science & Applications **8**, 90 (2019).
2. H.-L. Zhou, D.-Z. Fu, J.-J. Dong et al., Light: Science & Applications **6**, e16251-e16251 (2017).
3. A. D'Errico, R. D'Amelio, B. Piccirillo et al., Optica **4**, 1350-1357 (2017).
4. Z. Lin, J. Hu, Y. Chen et al., Advanced Photonics **5**, 036006 (2023).
5. S. Li, P. Zhao, X. Feng et al., Opt. Lett. **43**, 4607-4610 (2018).
6. S. Zhao, S. Chen, X. Wang et al., Opt. Lett. **45**, 5990-5993 (2020).
7. H. Wang, Z. Zhan, F. Hu et al., PhotoniX **4**, 9 (2023).
8. J. Zhu, Y. Wu, L. Wang et al., Physical Review Applied **20**, 014010 (2023).
9. A. V. Oppenheim, and R. W. Schafer, *Discrete-time signal processing* (Prentice Hall Press, 2009).